# Coarse-graining collective skyrmion dynamics in confined geometries

Thomas Brian Winkler[1*], Jan Rothörl[1], Maarten A. Brems[1], Hans Fangohr[2,3*], Mathias Kläui[1]

1 - Institute of Physics, Johannes Gutenberg University, Staudinger Weg 7, 55128 Mainz, Germany

2 - Faculty of Engineering and Physical Sciences, University of Southampton, University Road, Southampton SO17 1BJ, United Kingdom

3 - Max Planck Institute for the Structure and Dynamics of Matter Hamburg, Luruper Chaussee 149, 22761 Hamburg

\* Email: twinkler@uni-mainz.de, klaeui@uni-mainz.de

## Abstract

Magnetic skyrmions are magnetic quasi-particles with enhanced stability and different manipulation mechanisms using external fields and currents making them promising candidates for future applications for instance in neuromorphic computing. Recently, several measurements and simulations have shown that thermally activated skyrmions in confined geometries, as they are necessary for device applications, arrange themselves predominantly based on commensurability effects. In this simulational study, based on the Thiele model, we investigate the enhanced dynamics and degenerate non-equilibrium steady state of a system in which the intrinsic skyrmion-skyrmion and skyrmion-boundary interaction compete with thermal fluctuations as well as current-induced spin-orbit torques. The investigated system is a triangular-shaped confinement geometry hosting four skyrmions, where we inject spin-polarized currents between two corners of the structure. We coarse-grain the skyrmion states in the system to analyze the intricacies of skyrmion arrangements of the skyrmion ensemble. In the context of neuromorphic computing, such methods address the key challenge of optimizing read-out positions in confined geometries and form the basis to understand collective skyrmion dynamics in systems with competing interactions on different scales.

# Introduction

Magnetic skyrmions [1] are of major interest for next-generation spintronic applications due to their topological stabilization and a multitude of mechanisms for controlled nucleation, movement and annihilation. In previous magnetic skyrmion research, the focus has been mainly on using skyrmions for data storage devices, such as the skyrmion racetrack [2]. Recently additional schemes for using skyrmions for non-conventional computing, such as Brownian-based token computing [3], probabilistic computing [4] or Reservoir computing [5], have been developed.

The latter scheme exploits the non-linear nature of physical systems to perform calculations. The input data for the reservoir is transformed into an excitation of the physical system which introduces a complex non-linear response altering the system's high-dimensional phase space state. In this high-dimensional space, classification of the input data is usually significantly more straightforward, and the input separability is often conserved when extracting measurable quantities from the system. Simulation studies have already demonstrated that a skyrmion reservoir exhibits enough memory capacity, computing capacity as well as non-linearity to classify signals [6]. The highest score for the TI-46 benchmarking set for audio signals has been reached with a skyrmion reservoir [7].

The first skyrmion reservoir computer has been experimentally realized [8]. In this work it has been demonstrated that a single skyrmion in a triangular confinement suffices to realize 2- and 3-input Boolean operations including the non-linearly-separable XOR [8]. A triangular confinement was chosen due to its simplicity and ability to contact the edges with electric potentials. Instead of a skyrmion ensemble only a single skyrmion has been chosen to be hosted in the confinement, since skyrmions also exhibit non-linear behavior in various aspects. For example, non-linear interaction potentials between skyrmions and boundaries (as well as between different skyrmions) have been found [9], as well as other non-linear responses [10–13]. First studies have shown the possibility to perform different kinds of Boolean logic operations with this device [8]. Building on this, one can insert multiple skyrmions into the confinement to potentially increase the computing capacity of such a device.

Recent experimental and numerical studies of collective behavior of skyrmions in confined geometries has shown that diffusion of skyrmions heavily depends on the ability of the skyrmions to arrange themselves in patterns that are commensurate with the sample geometry [14]. For triangular confinements in particular, arrangements of 3, 6 and 10 skyrmions lower the average diffusion of the skyrmions, since those ensembles can arrange in a lattice that fills most of the space in the confinement [14,15].

However, independent of the commensurability effect, skyrmions arrange in certain ways in a confinement, due to the interplay of repulsive skyrmion-skyrmion and skyrmion-boundary interactions [16]. All the above-mentioned studies have been performed in the absence of external forces except for thermal fluctuations. We take the next step beyond these investigations by using computer simulation studies, in which we investigate a thermally activated 4-skyrmion system in a triangular confinement. In this system skyrmion interactions and thermal effects compete with current-induced dynamics. Currents are applied by applying an electric potential at contacts at the edges of the confinement. Current-induced spin-orbit torques (SOT) are employed as they are a promising way to encode input signals for reservoir computing schemes. We analyze the system's behavior by considering the different prominent states. We employ coarse-graining methods to determine those non-equilibrium steady states (NESS) based on Markovian processes and find via coarse-graining methods that several degenerate non-equilibrium states are possible when external forces are applied. By this we demonstrate that coarse-graining approaches are very useful to identify multiple energetically similar or even degenerate non-equilibrium steady states in driven systems with thermal excitations, as present for instance in Brownian reservoir computing.

# Methods

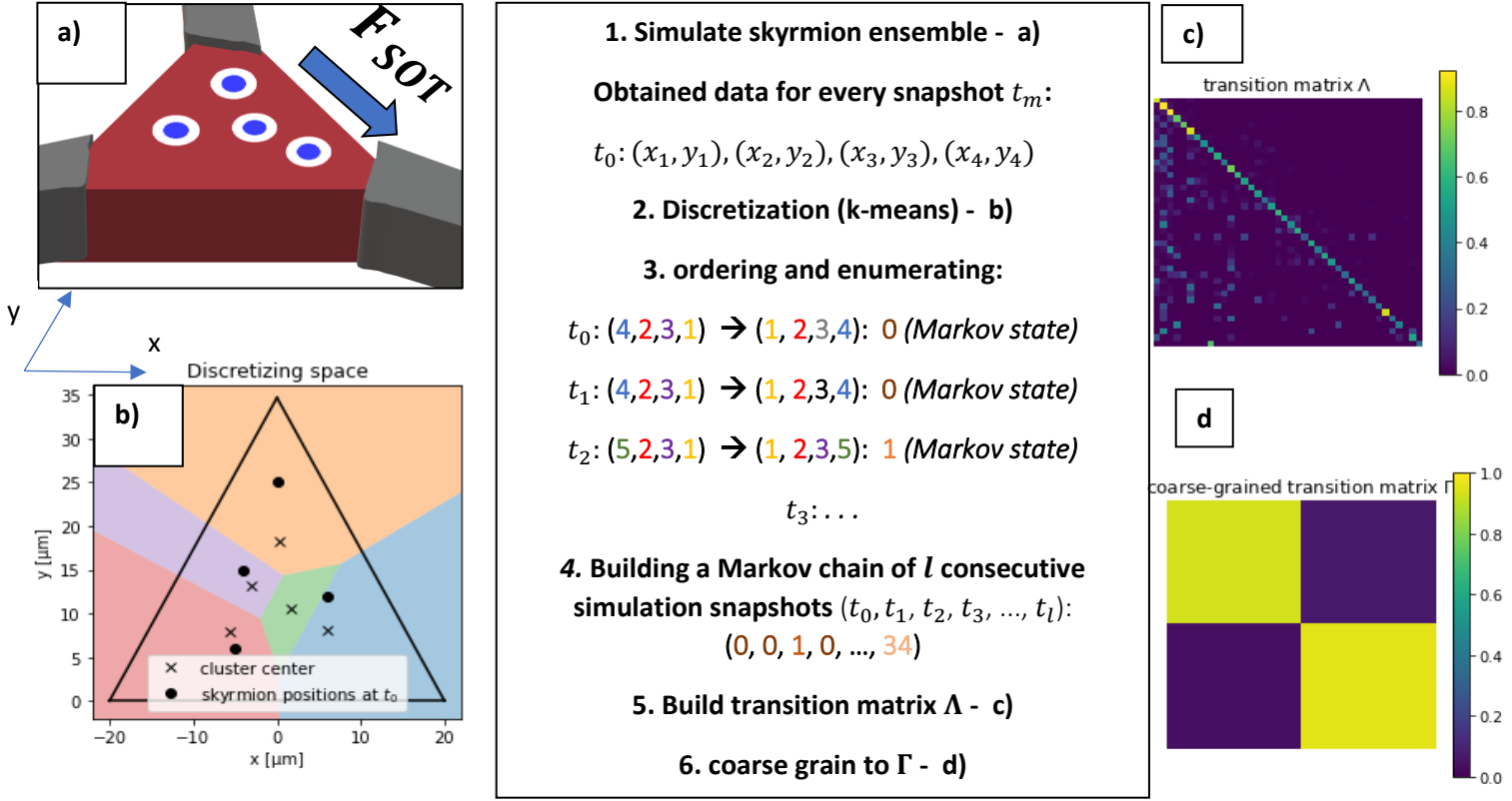

Figure 1: Visualizing the data analysis from simulation to coarse-graining. a) Schematic representation of the triangular confinement hosting four skyrmions. The overall direction of the force generated by the spin orbit torque is indicated. The current density distribution in the device was simulated in detail numerically. A similar setup with one skyrmion was used in Ref. [9]. b) For every used simulation snapshot the space was discretized with the k-means++ algorithm. For clarity an arbitrary skyrmion arrangement and only $k = 5$ cluster centers are chosen here. The discretized data is then converted into a Markovian process, see the center box. c) From this Markov process we can build the transition matrix $\Lambda$ and run the GPCCA algorithm to obtain d) a coarse-grained transition matrix $\Gamma$.

Simulations are performed in a rigid-particle based ansatz within the Thiele-approach [17,18]. We employ skyrmion interaction potentials extracted for a Ta/CoFeB/Ta/MgO/Ta thin film multilayer stack in [16]. We consider an equilateral triangle with side length $a = 40$ µm. The total force $F(r)$ acting on every skyrmion can be split up into internal force $F_{\text{int}}(r)$ and external forces $F_{\text{ext}}(r)$,

$F(r) = F_{int}(v) + F_{ext}(r)$, with $r$ the position of the skyrmion and $v$ its velocity. Internal forces are the dissipative force $F_G(v)$ and the gyroscopic force $F_G(v)$. External forces are the thermal force $F_{th}$, the skyrmion-skyrmion interaction force $F_{SkSk}(r)$ the skyrmion-boundary interaction force $F_{SkB}(r)$ and the force that arises due to the spin orbit torque, $F_{SOT}(r)$. The timestep of the simulations was chosen as $\tau = 0.1$, and we integrated over $2 \cdot 10^6$ time steps. We model the spatial dependence of $F_{SOT}(r)$ by setting it to be proportional to the distribution of current density in the system. The latter is obtained using minimalistic COMSOL Multiphysics® simulations described in the supplementary material [19]. The first $2 \cdot 10^5$ steps are not considered for the analysis, as they comprise an initialization process. For our analysis, the position of each skyrmion at every time step is stored. Further, to describe a system as a Markovian process, we must discretize our space. Spatial discretization is performed using a k-means++ clustering [20] algorithm with $k = 20$ on an array of all skyrmion positions that occur during the simulation.

**Figure 1** shows the data evaluation protocol in detail. Every snapshot of the simulation can be described by an array of four cluster centers since every configuration is an ensemble of four skyrmions. We choose $k = 20$ for a good trade-off between discretization and accuracy, since the number of theoretically available micro-states are $n = k^4$. We are reducing the 4-dimensional (4D) state into a 1D state, enumerating the discretized skyrmion configurations according to their appearance. This list of states is considered to be a Markovian process, from which we can build a transition matrix $\Lambda^{n \times n}$, with $n$ the number of states appearing in the process. If two skyrmions switched their states, the physical systems stays the same, since skyrmions are undistinguishable particles. To delete these permutations in the states, we are sorting the 4D-states in ascending order. For the coarse graining into $n_{coarse}$ states, we employ the generalized Peron cluster cluster analysis (GPCCA), which uses Schur decomposition techniques to coarse-grain the system [21]. We choose this algorithm as it can be applied to dynamical systems that are driven out of equilibrium, contrary to a classical eigenvector analysis. GPCCA calculates a coarse-grained transition matrix $\Gamma^{n_{coarse} \times n_{coarse}}$.

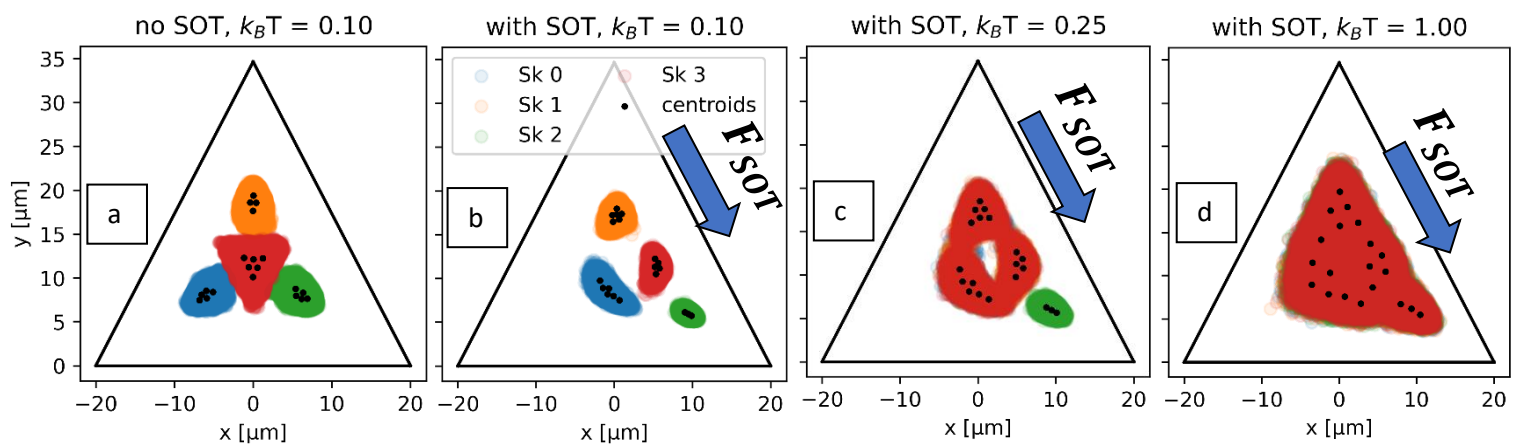

*Figure 2: Explored space of all skyrmions during the simulated time for the different temperature regimes. a) Low temperature regime without applied currents and thus zero SOT. Skyrmions try to arrange themselves according to the geometrical confinement [14]. The central skyrmion is showing the highest diffusion. b) Low temperature regime with injected currents: The skyrmions stay approximately at their position. c) Intermediate temperature regime, the 3 central skyrmions can change their position, while the bottom right skyrmion stays in its corner. d) High temperature regime: all skyrmions can explore the energy landscape of the device. Cluster centers of the k-means algorithm are plotted as black dots in all panels, which are used to discretize the system.*

## Results and Discussion

Our simulations reveal three different temperature regimes for the system's dynamics. **Figure 2** shows scatter plots of all the positions a skyrmions has been found during the simulation at different temperatures. We observe a low temperature regime, where each of the skyrmion stays approximately at its position with nearly no overlap between the probability density functions (PDF) of the skyrmions (Fig. 2a). There is an intermediate temperature regime, where the 3 skyrmions which are closest to the center can interchange positions (nearly identical PDFs), but the skyrmion close to the bottom-right corner of the device does not move significantly (Fig. 2 b-c). Finally, there is a high temperature regime, where all 4 skyrmions in the confinement have sufficient thermal energy to exchange positions and explore the whole accessible space (Fig. 2 d). The current is injected from the bottom right contact to the top contact, so the skyrmions are pushed with the electron flow to the bottom right corner.

To build the transition matrix $\Lambda^{n\times n}$, we assign every skyrmion at every simulation snapshot to the closest cluster and sort the arrangement in ascending order. Thereby, we reduce the 8-dimensional vector (x- and y- coordinates of each of the 4 skyrmions) to a 4-dimensional assignment. By sorting we also remove possible permutations, which is desired as those states are not distinguishable from a physical point of view. To reduce the assignment to a Markovian process, we just number the 4-dimensional assignment depending on its first appearance to obtain a Markov chain. We call $n$ the number of explored states and build the transition $\Lambda^{n\times n}$. The GPCCA algorithm can now be applied to $\Lambda^{n\times n}$ as visualized and described in **Figure 1**. We scan $n_{coarse}$ in a reasonable range and plot in **Figure 3** the quality measures provided by the GPCCA method for the coarse-grained system.

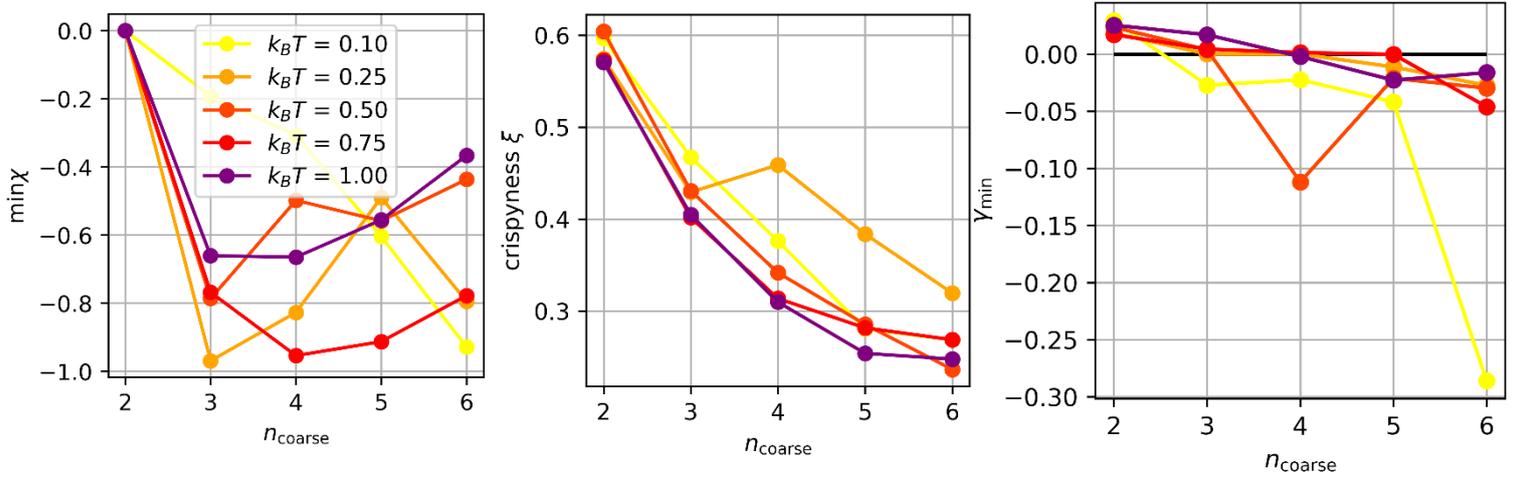

Figure 3: Quality-measures of the coarse-graining for different Temperatures. A small negative $min\chi$ indicates well-separable clusters. The crispiness $\xi$ indicates a long dwell time. A positive $\gamma_{min}$ ensures $\Gamma$ to describe a Markovian process. All measures indicate an optimal number $n_{coarse} = 2$ meaning that there are 2 predominant states present. For 3 states already negative entries in $\Gamma$ occur for different temperatures, indicating a failing of the algorithm for $n_{coarse} = 3$.

A high crispiness $\xi$, indicating a long dwell time in each coarse state on average, as well as a $min\chi$ close to zero, but negative, indicates well-separable clusters. Further we check the smallest element $\gamma_{min}$ in the coarse-grained transition matrix $\Gamma^{n_{coarse}\times n_{coarse}}$. Non-negative values in $\Gamma$ are a requirement for interpreting the matrix as a transition matrix for a Markovian process. For details on the algorithm, we refer to Ref. [21].

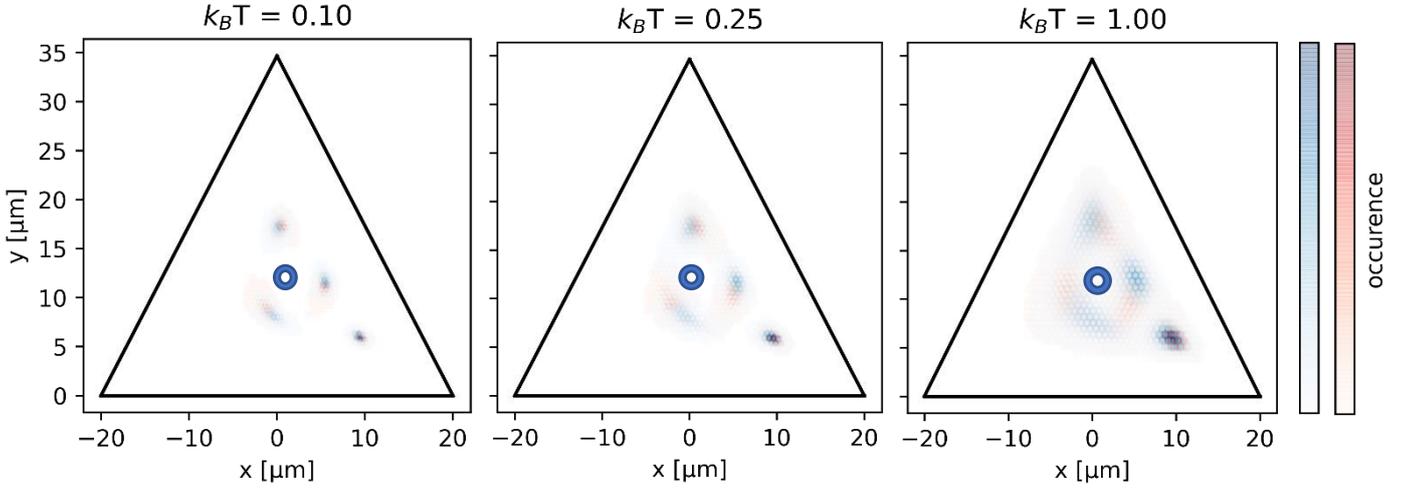

*Figure 4: Back-mapping of the coarse-grained clusters to the unscaled probability of a skyrmion located in the confinement. Red probability clouds refer to the "arrow" state, blue ones to the "rhombus" state. For small temperatures the peaks of both states are close together, while for higher temperatures the probability clouds broaden and the states get more distinct. The approximate rotation center of the three central skyrmions is drawn with a blue ring.*

**Figure 3** shows $min\chi$, $\xi$ and $\gamma_{min}$ for a various number of $n_{coarse}$. The measures indicate an optimal number of coarse-grained states of $n_{coarse} = 2$. It shows the highest crispiness, a small appropriate $min\chi$ and a small positive $\gamma_{min}$ value. Even for 3 clusters, $\gamma_{min}$ is not positive for all temperatures. For 5 and more clusters, no coarse-grained model can be created with the GPCCA method. This means that 2 states can be distinguished that predominantly occur.

**Figure 4** depicts the occurrence maps for the skyrmions being in one of the identified metastable states. We observe that introducing spin torques into the system leads to a shifting of the pdf for all skyrmions. For applied spin torques we find two configurations with the GPCCA approach, of which one arrangement looks like an "arrow" (red) and one looks like a "rhombus" (blue). Both states have one skyrmion trapped in the bottom right corner, while the three skyrmions closer to the center rotate around each other, driven by SOT. In the arrow state, one of the central skyrmions is closer to the bottom right one, leading to a slight shift in the probability density of the latter one towards the corner. The splitting of the two states occurs due to the interplay of SOT and the repulsive potential of each skyrmion and the boundary. In both configurations, the average skyrmion distance to its closest neighbors is approximately constant, as visible in **Figure 4**. Furthermore, the two states become more distinct with increasing temperature.

We want to underline that this kind of analysis can be a key factor in understanding thermal skyrmion dynamics in confined geometries. It is useful to find optimal positions for read-out contacts, such as magnetic tunnel junctions [9]. Further, when the strength of the spin torque is changed, there might be an abrupt change in the variety of possible states of the system. Finally, also the geometry of the confinement can be optimized using this approach.

## Conclusion

To conclude, we have expanded the analysis of the commensurability effects in skyrmion ensembles in confined geometries to systems that are driven out of equilibrium. We have shown that commonly used coarse-graining methods in statistical physics can be applied to the dynamics of skyrmion ensembles. In our case - four skyrmions in a triangular confinement and driven by SOT – we have identified two states with similar occurrence probability, namely an arrow-like and a rhombus-like state, while the arrow-like state occurs roughly 40 % of the time, the rhombus-like state roughly 60 % of the time. The states can transition into each other by rotating the three central skyrmions by some degree. The dynamics of skyrmion ensembles in confined geometries is a key factor to understand the underlying physics and to develop neuromorphic computing devices based on skyrmions, such as Brownian Reservoir computing devices. In particular, the presented methods allow one to optimize the confinement shape and the position of read-out contacts, such as magnetic tunnel junctions. Thereby, these methods may be a key tool for studies to enhance the applicability, realizability, performance and efficiency of non-conventional computing methods.

# Authors Declaration

## Acknowledgment

The authors acknowledge funding from the emergentAI center, funded by the Carl Zeiss Stiftung, as well as by the German Research Foundation (DFG SFB TRR 173, SPIN+X, A01 - 268565370 - and B12 – 268565370) and project 403502522 SPP 2137 as well as TopDyn. This project has received funding from the European Research Council (ERC) under the European Union's Horizon 2020 research and innovation programme under grant agreement No. 856538 (ERC-SyG 3D MAGIC). M.B. is supported by a doctoral scholarship of the Studienstiftung des deutschen Volkes. We thank Peter Virnau for valuable discussions.

## Authors contribution

T.B.W. performed the data analysis, the coarse graining and wrote the initial draft. J.R. performed particle-based simulations of skyrmion ensembles. M.B. performed simulations of current density distributions in the confinement. M.K. and H.F. supervised the study. All authors contributed to the manuscript.

# Data availability

The data and codes that support the findings of this study are available from the corresponding authors upon reasonable request.